\documentclass[letterpaper, 10 pt, conference]{ieeeconf}  % Comment this line out
\IEEEoverridecommandlockouts                              % This command is only
\usepackage{lineno} % To obtain line numbers
\usepackage{bm}

\usepackage{amssymb}
\usepackage[ruled,vlined]{algorithm2e}
\usepackage{cite}
\usepackage{graphicx}
\usepackage{amsmath}
\usepackage{array}
\usepackage{stfloats}
\usepackage{url}
\usepackage{xcolor}
\usepackage{comment}
\usepackage[hidelinks]{hyperref}

\title{\LARGE \bf
Activity Detection And Modeling Using Smart Meter Data: \\Concept And Case Studies
}

\author{Hao Wang, Gonzague Henri, Chin-Woo Tan, Ram Rajagopal
\thanks{Hao Wang, Chin-Woo Tan, and Ram Rajagopal are with the Department of Civil and Environmental Engineering, Stanford University, CA 94305, USA
        {\tt\small \{haowang6,tancw,ramr\}@stanford.edu}}
\thanks{Gonzague Henri is with Stanford University, CA 94305, USA and Total, R\&D, USA
        {\tt\small gonzague.henri@total.com}}
}
\begin{document}
\maketitle
\thispagestyle{empty}
\pagestyle{empty}

%%%%%%%%%%%%%%%%%%%%%%%%%%%%%%%%%%%%%%%%%%%%%%%%%%%%%%%%%%%%%%%%%%%%%%%%%%%%%%%%
\begin{abstract}
	
	Electricity consumed by residential consumers counts for a significant part of global electricity consumption and utility companies can collect high-resolution load data thanks to the widely deployed advanced metering infrastructure. There has been a growing research interest toward appliance load disaggregation via nonintrusive load monitoring. As the electricity consumption of appliances is directly associated with the activities of consumers, this paper proposes a new and more effective approach, i.e., activity disaggregation. We present the concept of activity disaggregation and discuss its advantage over traditional appliance load disaggregation. We develop a framework by leverage machine learning for activity detection based on residential load data and features. We show through numerical case studies to demonstrate the effectiveness of the activity detection method and analyze consumer behaviors by time-dependent activity modeling. Last but not least, we discuss some potential use cases that can benefit from activity disaggregation and some future research directions.
	
\end{abstract}

%%%%%%%%%%%%%%%%%%%%%%%%%%%%%%%%%%%%%%%%%%%%%%%%%%%%%%%%%%%%%%%%%%%%%%%%%%%%%%%%
\section{Introduction} \label{intro}

\subsection{Background}
Thanks to the widespread deployment of smart meters, high volumes of residential load data have been collected and made available to both consumers and utility companies. In the United States (US) alone, more than 65 million smart meters have been installed by 2015 and the number of installations is expected to be 90 million by 2020 \cite{cooper2016electric}. %The massive volumes of load data cultivate great commercial potentials and data analytics market that is growing to 3.8 billion US dollars in 2020 per the estimation in \cite{fairchild2013soft}. 
For consumers, having access to the real-time load data help them better understand their load patterns and wisely plan the consumption to conserve energy and save electric bills, especially under a time-of-use tariff structure \cite{zhou2016understanding}. For utility companies, analyzing consumer load data enables better residential load forecasting and engagement of consumers in the planning and operation of the system, for example through demand response programs \cite{weng2018probabilistic}. Therefore, smart meter data analytics opens up new opportunities to both consumers and utility companies toward a smart grid.

Various analytical techniques have been developed to analyze smart meter data through machine learning \cite{zhang2018big}. For example, a long short-term memory model was used to forecast the load of individual residential households in the short term \cite{kong2017short}. An adaptive K-means method was proposed to identify representative load profiles %based on the load data of 220 thousand 
of consumers and characterize consumers' lifestyles \cite{kwac2016lifestyle}. A recent study in \cite{tang2019leveraging} took a further step and developed a deep neural network model to reveal the connection between consumers' load patterns and their socioeconomic characteristics including age, income, and educational level. But the aforementioned studies in \cite{kong2017short,kwac2016lifestyle,tang2019leveraging}, despite their applications (e.g., load forecasting, load segmentation, and load pattern prediction), all focused on the aggregated load of consumers. However, appliances play a significant role in the total electricity consumption of residential consumers \cite{huebner2016understanding}. Kavousian et al. in \cite{kavousian2013determinants} found that specific appliances like refrigerators and entertainment devices are highly associated with greater daily minimum electricity load, and in contrast, electric water heaters and dryers contribute significantly to daily maximum consumption. To encourage consumers to change consumption behaviors and save energy, it is essential to disaggregate and monitor electricity consumption at the appliance level. 

Several load disaggregation approaches have been developed to provide detailed energy consumption of consumers. To this end, a load disaggregation algorithm was developed in \cite{koutitas2015low} to decompose smart meter data into discrete pulses, each of which is associated with a registered appliance. % via maximum likelihood procedure.
The study in \cite{liu2015generic} measured energy consumption down to appliances (e.g., lighting, refrigerator, and microwave) to identify inefficient devices. Among different categories of disaggregation approaches, non-intrusive load monitoring (NILM) methods have attracted a lot of attention to disaggregate total load into the individual load of appliances \cite{hosseini2017non}. But traditional NILM for appliance load disaggregation did not achieve high accuracy and has come to a halt due to its inherent drawbacks. Some argued in \cite{hosseini2017non} that it is impractical to identify appliance load solely based on the measurement of load signal at a single point. Also, from the perspective of utility, it is challenging to engage consumers by solely providing split-up appliance consumption because there lacks interactive interpretation between consumers' behaviors and appliance consumption. Aiming at these problems, in this paper, we propose a more interactive and less complex approach toward load disaggregation, namely \emph{activity disaggregation}. Detecting daily activities of consumers is promising to achieve higher accuracy and gain insights into consumer behaviors, based on which utilities and consumers can improve their operations and satisfaction, respectively.

\subsection{Main Results}

This paper proposes the use of smart grid data to unveil consumer behaviors in terms of activities. Given the exploratory nature of the work, we present the idea of activity disaggregation and develop a framework for activity detection. Using realistic load data from Pecan Street \cite{street2016pecan}, we show some numerical results on activity detection and modeling, which shed light on analyzing consumer behaviors. Specifically, we summarize the main body of this paper as follows. 
\begin{itemize}
	\item In Section \ref{overview}, we review existing studies on load disaggregation and analyze its major drawbacks that motivate us to study activity disaggregation. We present the concept of activity detection and discuss its advantages over traditional appliance load detection and disaggregation.
	\item In Section \ref{activity}, we develop a framework that leverages multiple data resources, feature engineering, and machine learning techniques for activity detection. We also discuss activity modeling approaches to the understanding of consumer behavior.
	\item In Section \ref{case}, we take house-cooling as an example to validate our proposed activity detection framework using multiple features in both time and frequency domains. We also show the activity modeling based on realistic data and discuss potential use cases for consumers and the utility. 
	\item In Section \ref{conclu}, we summarize this work and discuss some future research directions.
\end{itemize}

\section{Overview of Load Disaggregation and Motivation} \label{overview}
We review related work on load disaggregation and discuss our motivation toward activity disaggregation.

\subsection{Related Works}
Firstly proposed by Hart in \cite{hart1992nonintrusive}, NILM aims to find the energy consumption for individual appliances and is essentially a single channel blind source separation problem. As a highly underdetermined problem, it is challenging to decompose the aggregated load into many appliance loads. In the past decade, many NILM methods have been proposed for residential consumers and can be classified into two categories \cite{henriet2018generative}. The first category of NILM methods \cite{barsim2015toward,girmay2016simple} rely on event detection, for example, on and off transition of appliances and change of operational modes or states of appliances. However, there is an underlying assumption that the load change is caused by only one appliance, but such an assumption often does not hold in practice. The second category of NILM methods in \cite{kolter2012approximate,rahimpour2017non} assumes that the aggregated load is a mixture of a number of unknown load signals, each of which is associated with an appliance. Therefore, the second category method aims to recover the detailed split-up appliance consumption. However, their performance is highly dependent on the sampling frequency and smart meters often do not support such high-frequency data sampling and storage \cite{dong2014fundamental}. For more related works, readers can be referred to the survey papers in \cite{hosseini2017non, tabatabaei2016toward}.

\subsection{Motivation}
Although some existing NILM approaches could potentially disaggregate some appliances in specific settings and improve energy efficiency, the existing studies often suffer from 
\begin{itemize}
	\item low disaggregation accuracy,
	\item and difficulty in interpreting appliance use and engaging consumers.
\end{itemize}
and thus motivate our study of activity detection. 

\begin{figure}[!htbp]
	\centering
	\vspace{-1mm}
	\includegraphics[width=.35\textwidth]{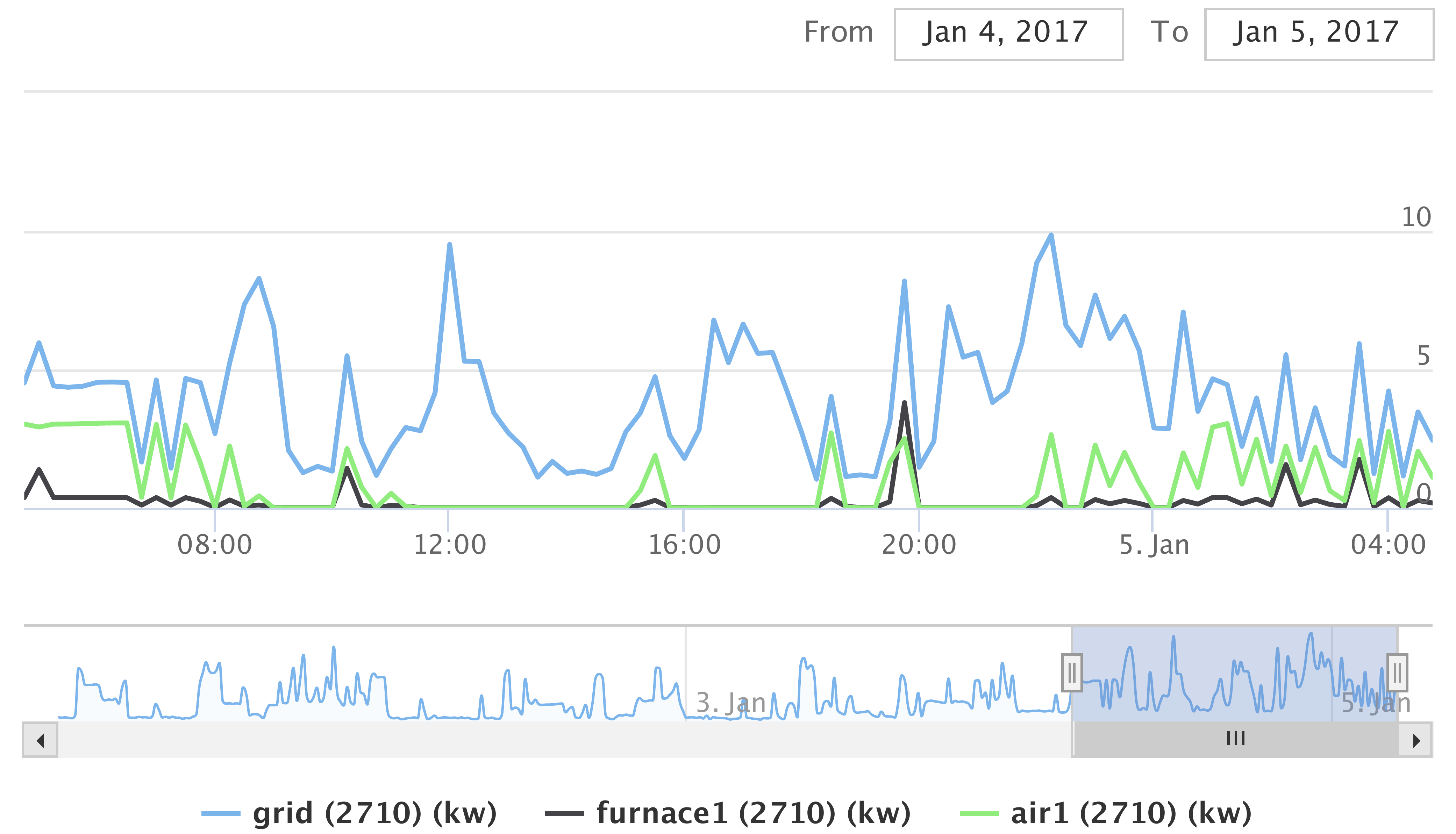}
	\vspace{-2mm}
	\caption{Energy consumption of furnace and air conditioning.}\label{heating}
	\vspace{-1mm}
\end{figure}

For example, using Pecan street load data, we show in Fig. \ref{heating} that the energy consumption of furnace (air handler) and air compressor are perfectly matched in time but their consumption is different from time to time. This suggests that the load changes are associated with both furnace and air compressor and thus the aforementioned disaggregation methods can easily fail in this case. In fact, both furnace and air compressor reflect the same activity of house heating and thus we are motivated to study activity disaggregation instead of separated appliance load disaggregation.

The idea of activity recognition has been studied in the healthcare area to detect emergency situations \cite{avci2010activity} by monitoring daily activities and movements of patients. But such recognition techniques often rely on extra sensor devices and thus are intrusive and can be costly. Other studies, e.g., \cite{widen2010high}, built probabilistic models of activities that are associated with energy consumption. But the consumer behaviors are so diverse and there lacks a generic model to work for personalized consumption behaviors. Different from existing works, we restrict our work to be non-intrusive but aim to take full advantage of different features of load signals (e.g., in the time domain and frequency domain) and external information (e.g., weather) to detect activities and model activities to benefit utility service and consumer experience.

\section{Activity Detection and Modeling} \label{activity}
\begin{figure}[t] %[!htbp]
	\centering
	\includegraphics[width=.36\textwidth]{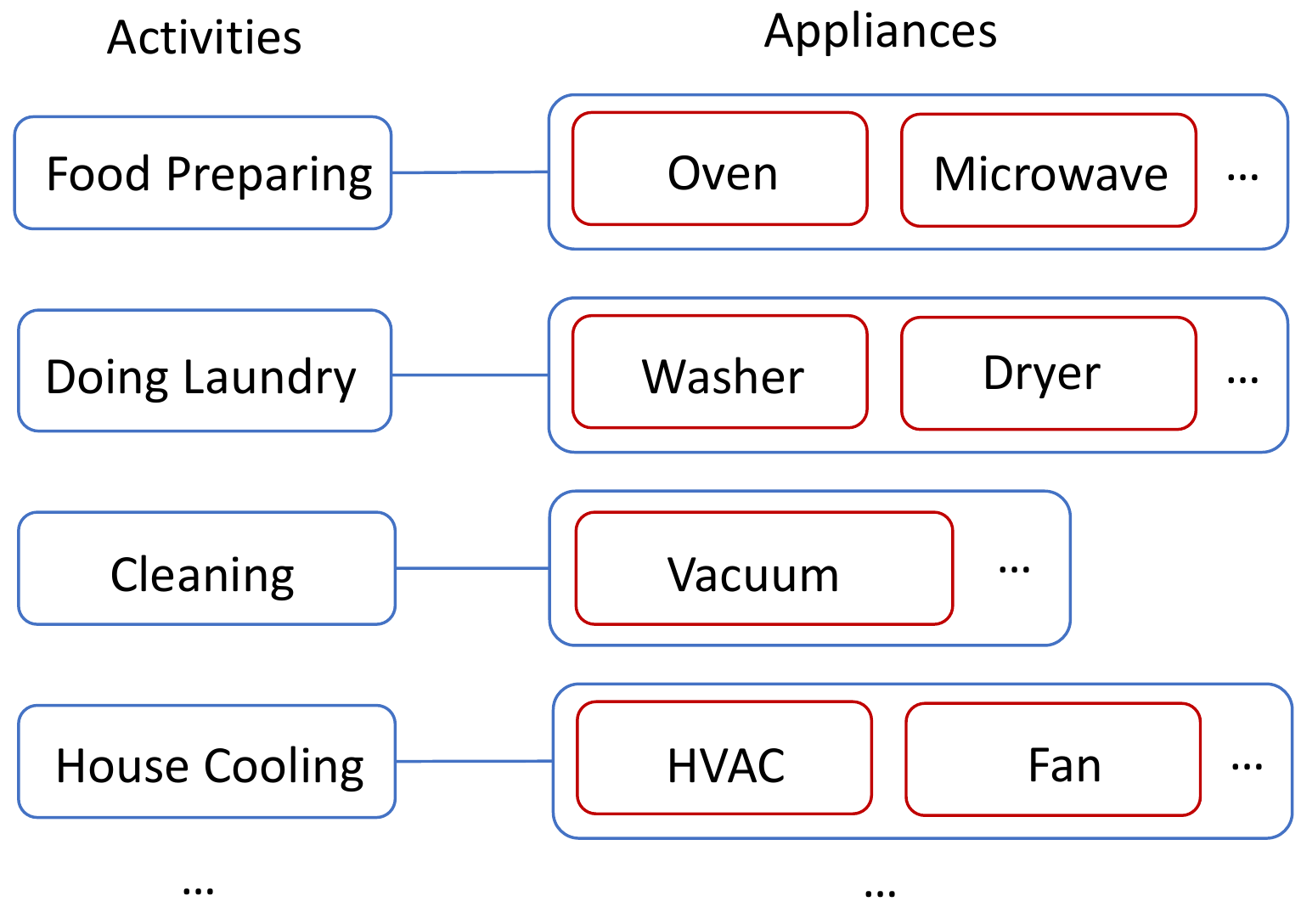}
	\vspace{-3mm}
	\caption{Activities and associated appliances.}\label{figure_activity}
	\vspace{-5mm}
\end{figure}

The energy consumption of different appliances is often associated with certain activities such as food preparation, relaxation, and cleaning. Our aim is to discover consumer activities using smart meter data. As shown in Fig. \ref{figure_activity}, doing each activity can involve the use of multiple appliances. When preparing food, consumers may use oven and/or microwave. For laundry, it is typical to operate cloth-washer and then a dryer. We will present a framework to extract features from load data and external information (e.g., weather) and detect activities. Then we present different activity modeling approaches. For example, activities such as grooming, food preparing, and doing laundry are usually regular routines. We will develop time-dependent modeling to illustrate detected activities to get insights into consumer behaviors.

\subsection{Activity Detection}
\begin{figure}[b] %[!htbp]
	\centering
	\vspace{-3mm}
	\includegraphics[width=.36\textwidth]{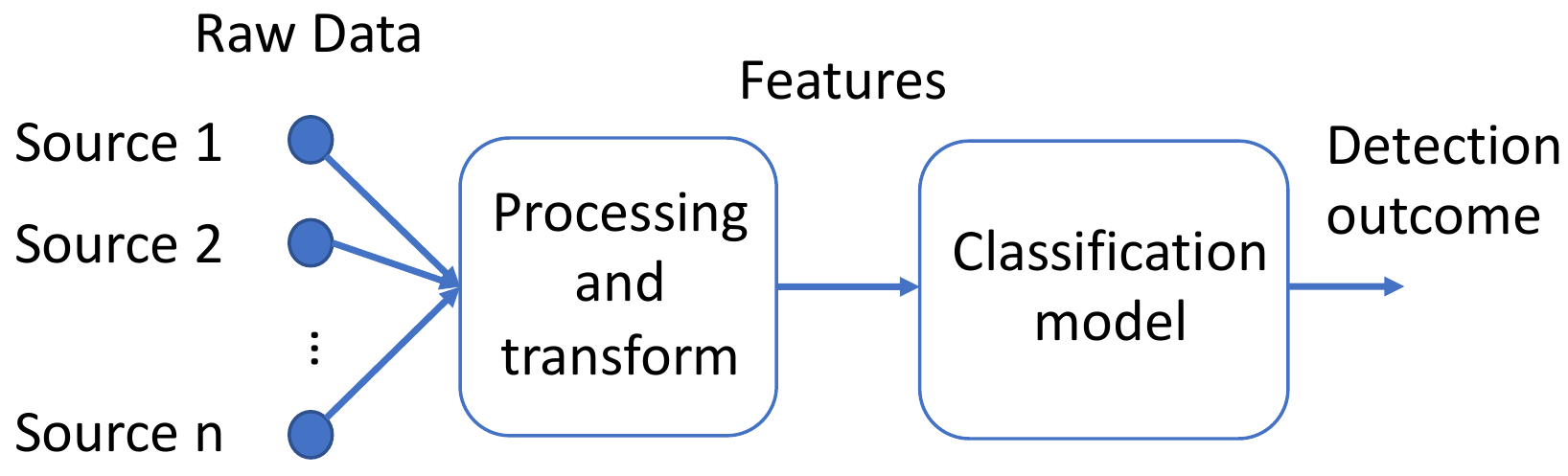}
	\vspace{-3mm}
	\caption{Activity detection model.}\label{figure_model}
\end{figure}

Fig. \ref{figure_model} shows the overview of the framework for activity detection, which consists of two major parts: feature extraction and classification model. Note that different activities may demand different features and models for effective detection, therefore, we do not specify the detection method in this section but propose the general framework. We will show a case study in Section \ref{case} to validate this framework for a specific task. 

The activity detection algorithm receives the data from the smart meters and other sources (e.g., weather). Note that the smart meter data can have different resolutions (e.g., 15 minutes, 1 minute, and 1 second) depending on the setup. We primarily rely on smart meter data, but also try to include other available information like temperature, which is proven to be correlated with energy consumption. When processing the load data, we aim to extract effective features as the input of the classification model. Those features can be the first-order derivative of total load and the variance of the load in a time interval. We call them time-domain features and they are usually good indicators of timing and energy intensity of activities. Besides, load patterns can also be captured by frequency analysis, e.g., fast Fourier transform (FFT). FFT can convert the load signal in the time domain to a representation in the frequency domain and thus provide new insights that are invisible in the time domain. Having a set of features extracted, we will choose and train a classification model and such a model can be based on logistic regression, random forest, support vector machine (SVM), and deep neural networks \cite{mohri2018foundations}, depending on the specific detection tasks.\footnote{It's worth mentioning that in a practical setting, the detection algorithm is expected to work on a set of activities of a large number of consumers but the available training data can be limited. Therefore, transfer learning \cite{pan2009survey} can be a promising technique to transfer knowledge across tasks (of activity detection) and improve the effectiveness of the detection algorithm. We will consider transfer-learning-based techniques in our future work.}

\subsection{Activity Modeling}
After detecting residential activities from load data, we need to create models to analyze activities and generate insights. There has been a number of models attempting to analyze customers' behaviors, e.g., in \cite{kwac2016lifestyle}. For example, Kwac et al. in \cite{kwac2016lifestyle} developed a clustering-based method to segment customers' lifestyles based on their load data, but such lifestyles do not differentiate customers' activities.

We introduce two approaches to activity modeling. One approach is to model the relationship between a set of activity and the time (e.g., hours of a day and different seasons) with a view to understanding how different activities exhibit time-dependent characteristics. Specifically, time dependence \cite{torriti2017understanding} refers to the high occurrence of the same activity over the same periods (e.g., hours of a day). From the perspective of behavioral science, understanding the timing of activities helps the study of social practices of consumers in terms of the ordering and overlapping over time. From the engineering perspective, understanding the timing of activities can contribute to load shifting and demand management. System peak load is a result of high consumption from activities at the aggregated level and we can identify the critical activities that contribute most to the system peak and then design proper demand management schemes to mitigate the problem.  

\begin{figure}[t] %[!htbp]
	\centering
	\includegraphics[width=.34\textwidth]{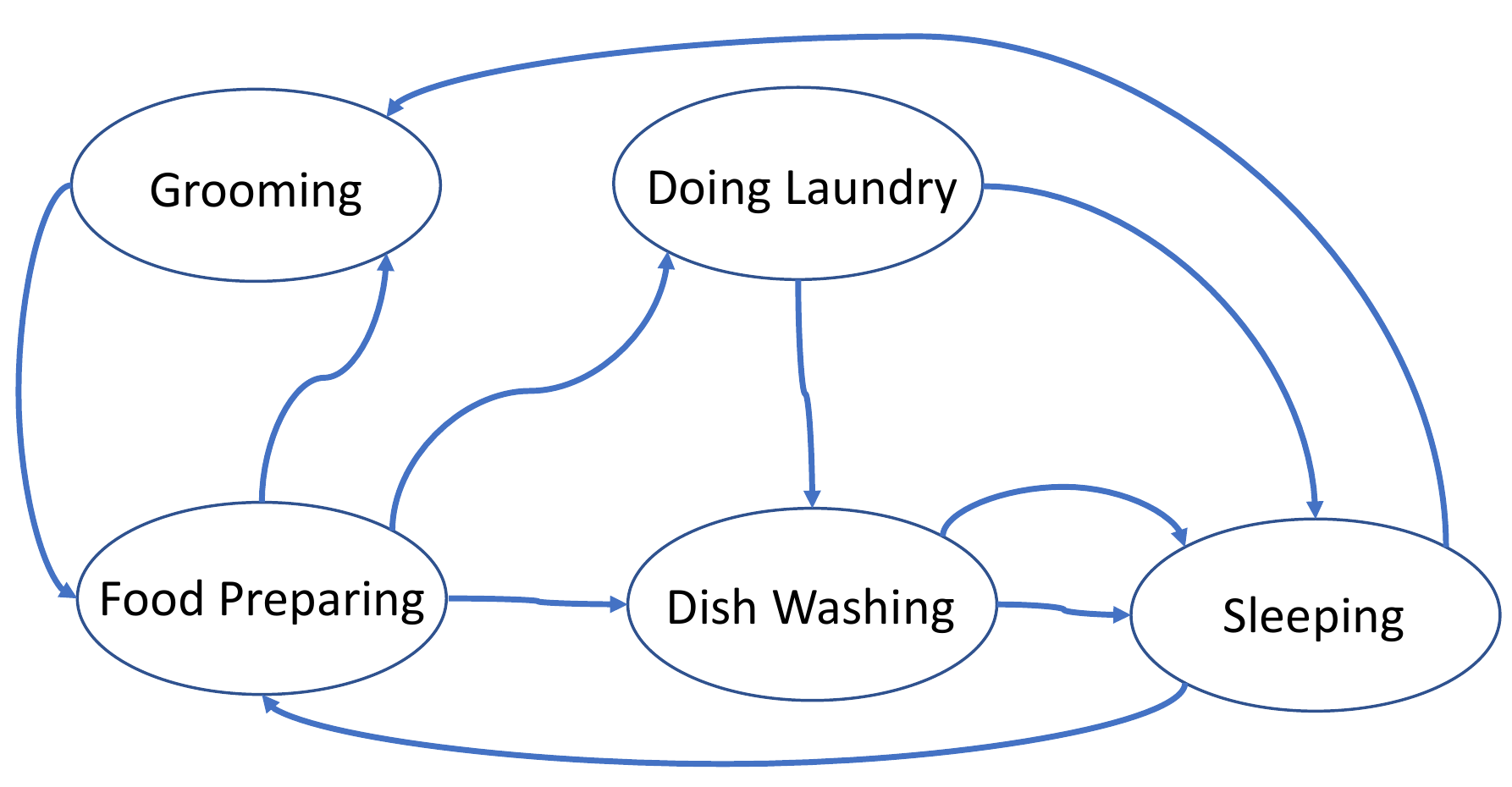}
	\vspace{-3mm}
	\caption{Activity transition.}\label{figure_transition}
	\vspace{-5mm}
\end{figure}
The other approach is to model the relationship among different activities based on their timing sequence. Social practice theory suggests that consumers' activities should be treated in a holistic manner to reflect the patterns, e.g., daily routines \cite{torriti2017understanding}. Understanding how consumers' routines are embedded can provide useful insights into the interpretation and predictability of everyday-life activities. To this end, the sequences of detected activities can be constructed as Markov-chain-like transitions of states (i.e., activities) as shown in Fig. \ref{figure_transition}. We take five activities (including sleeping, grooming, food preparing, doing laundry, and dish washing) as an illustrative example. Each arrow indicates a transition from one activity to another and is associated with a probability showing the likelihood of transition. We will numerically present a case using realistic load data in Section \ref{case}.

\section{Case Studies} \label{case}
In this section, we will present case studies based on realistic load data from Pecan Street \cite{street2016pecan}. We acquire 1-minute load data of 160 consumers with their load of appliances in 2017 and bundle appliances as the corresponding activities according to Fig. \ref{figure_activity}. In the following, we will show some preliminary results of activity detection and modeling.  

\subsection{Activity Detection Case Study}
We take house cooling/heating as an example and tailor a detection algorithm following the framework in Fig. \ref{figure_model}. Specifically, we consider minute-level load data of consumers and the temperature profiles as the input. We extract features in both time domain and frequency domain (via FFT). The time-domain features consist of the first-order derivative, variance of minute-level load profiles in every hour. We take the first ten elements of the amplitude spectrum as frequency-domain features. For the classification model, we choose SVM.\footnote{To serve the illustrative purpose, we do not discuss the performance of different classification models but will consider in the future work.} We split the data into 70\% for training and the rest for validation and test. The goal is to identify whether there is house cooling/heating activity in each detection window (i.e., hour) solely based on the aggregated load and additionally available information (e.g., the temperature in this case study). We aim to show the effectiveness of different features in the detection performance by comparing four cases: Method 1 (using frequency-domain features alone), Method 2 (using time-domain features alone), Method 3 (using both frequency-domain and time-domain features but not temperature), and Method 4 (using all available features).

We evaluate the performance of the proposed algorithm by the following metrics: accuracy, precision, and recall. In the detection outcomes, we denote true positive, true negative, false positive, and false negative as TP, TN, FP, and FN, respectively. Then, accuracy, precision, and recall are calculated as
\begin{equation*}
\begin{aligned}
\text{Accuracy} &= \frac{\text{TP}+\text{TN}}{\text{TP}+\text{TN}+\text{FP}+\text{FN}}, \\
\text{Precision} &= \frac{\text{TP}}{\text{TP}+\text{FP}},\\
\text{Recall} &= \frac{\text{TP}}{\text{TP}+\text{FN}}.
\end{aligned}
\end{equation*}

We first show the performance metrics of the compared methods. The results show that only using frequency-domain features or time-domain features are less effective in detecting cooling operations, e.g., 55.3\% and 57.3\% of recall for Method 1 and Method 2, respectively. Mixing frequency-domain and time-domain features significantly improves the performance and adding temperature further boosts the performance to achieve 98.3\% of accuracy.
\begin{figure}[t] %[!htbp]
	\centering
	\includegraphics[width=.32\textwidth]{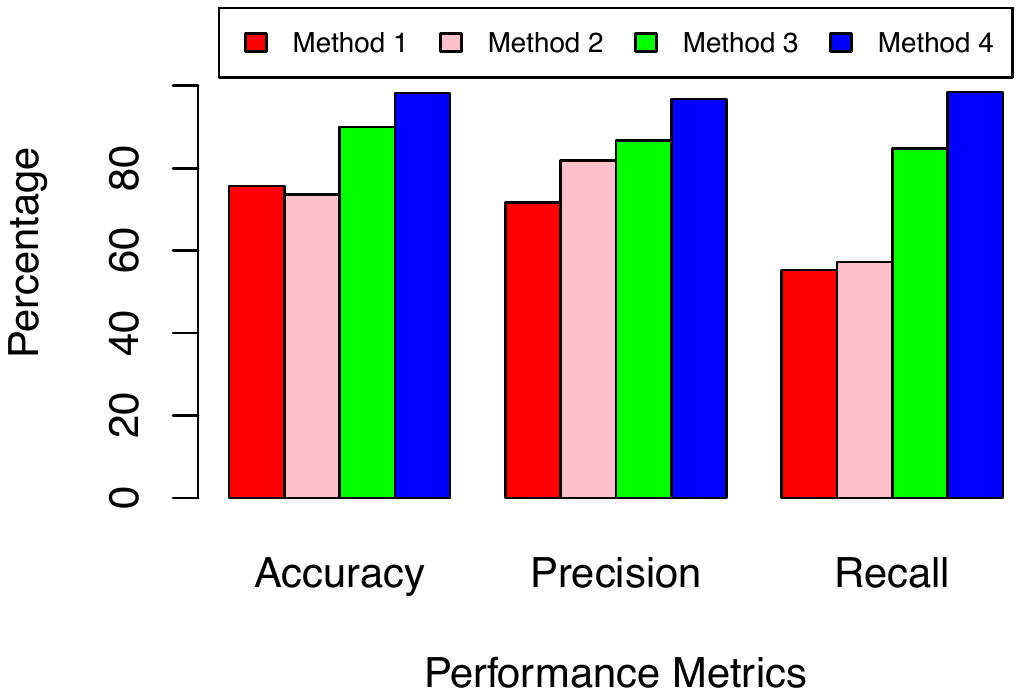}
	\vspace{-3mm}
	\caption{Performance comparisons  accuracy, precision, and recall.}\label{metrics}
	\vspace{-2mm}
\end{figure}

\begin{table}[!htbp]
	\centering
	\caption{Detection results of consumers (\#994 and \#871).} 
	{\begin{tabular}{|l|l|l|l|} 
			\hline
			Consumers & Accuracy & Precision & Recall \\ \hline \hline 
			\#994  & 99.40 & 100 & 99.10 \\ 
			\#871  & 99.40 & 100 & 99.32 \\ \hline
	\end{tabular}}{}
	\label{table_results}
	\vspace{-1mm}
\end{table}

\begin{figure}[!htbp]
	\centering
	\includegraphics[width=.36\textwidth]{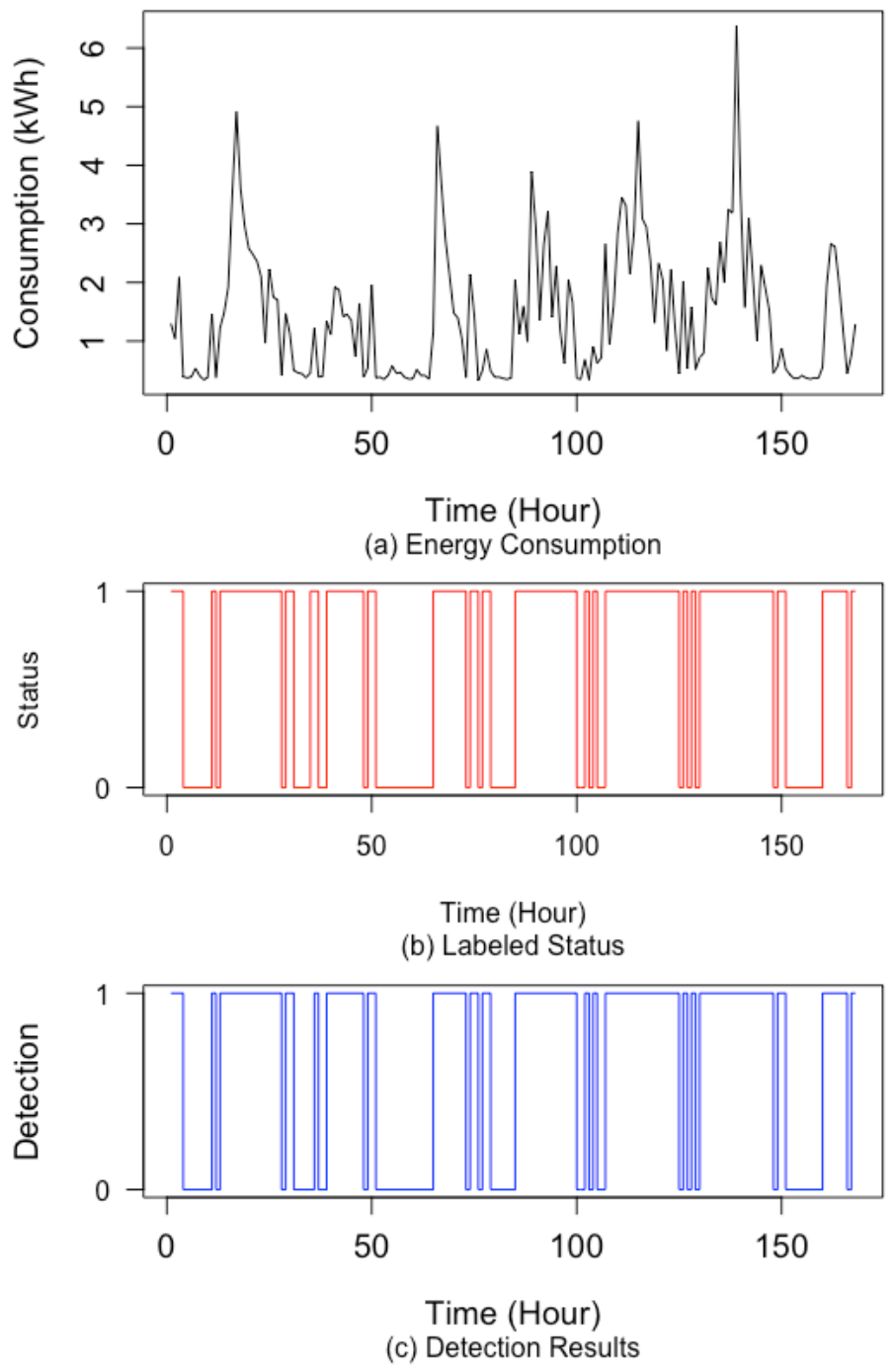}
	\vspace{-3mm}
	\caption{Energy consumption profiles, house-cooling ground truth, and detection results of consumer (\#994) in August 1-7, 2017.}\label{detection1}
	\vspace{-5mm}
\end{figure}
We also use the trained Method 4 to perform the detection on the aggregated load of two randomly selected consumers (\#994 and \#871) on August 1-7, 2017. Table \ref{table_results} shows the detection results and we see that Method 4 achieves high accuracy and perfect precision and there is only one FN in both tests. We further plot the aggregated load, ground-truth status, and detection results over time for consumers \#994 in Fig. \ref{detection1}. We see that the only one FN occurs at Hour 35 and the rest of detection results are accurate compared with the ground truth.

\subsection{Activity Modeling Case Study}
\begin{figure}[t] %[!htbp]
	\centering
	\includegraphics[width=.43\textwidth]{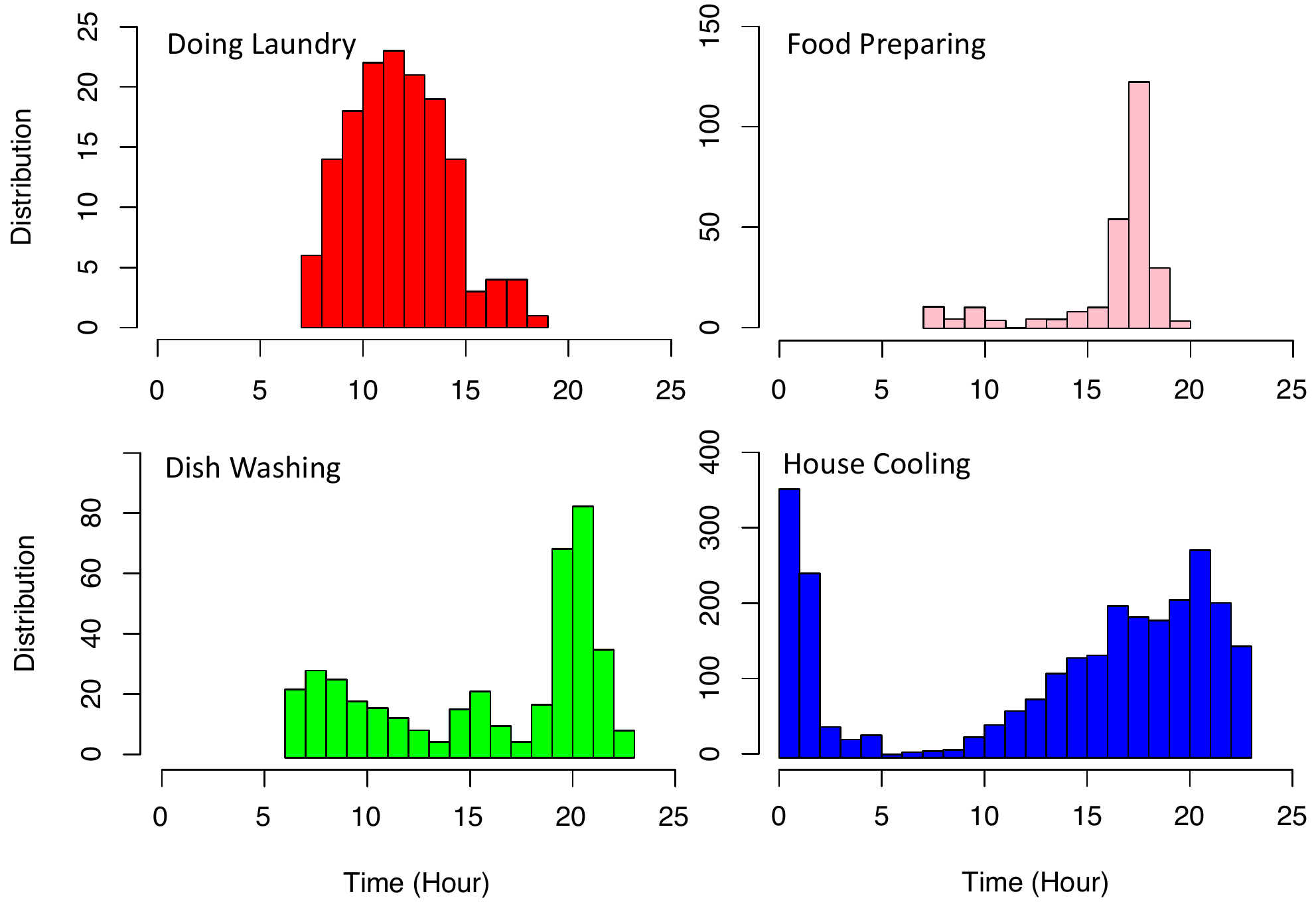}
	\vspace{-3mm}
	\caption{Distribution of time-dependent activities.}\label{distribution_activities}
	\vspace{-3mm}
\end{figure}

We also take Consumer \#994 as an example to numerically show the activity modeling. Fig. \ref{distribution_activities} depicts the distribution of time-dependent activities over 24 hours based on historical data. We see that food preparing and dish washing show similar time dependences and have the highest occurrence in the evening time. Therefore, they are identified as critical activities and proper intervention can help shave the system peak. Doing laundry and house cooling exhibit opposite patterns. Note that different consumers may exhibit very diverse distributions. Per the illustrative example in Fig. \ref{figure_activity}, we also calculate the transition matrix $P$ for Consumer \#1042 based on the historical data in May 2017 as follows. We define the activity state space as $\{$Sleeping, Grooming, Food-Preparing, Dish-Washing, Doing Laundry$\}$. From the transition matrix, we see that from sleeping, the residents have a higher probability to prepare food (i.e., breakfast) directly. The residents usually do the laundry and dish-washing before sleep. The showed activity modeling approaches can be helpful in understanding consumer behaviors and improve utility services. 
\begin{equation*}
P =
\begin{bmatrix}
0 & 1/6 & 5/6 & 0 & 0 \\
0 & 0 & 1/2 & 1/4 & 1/6 \\
1/12 & 5/12 & 3/10 & 1/5 & 0 \\
4/7 & 0 & 0 & 3/7 & 0 \\
0 & 0 & 0 & 1/2 & 1/2 \\
\end{bmatrix}
\end{equation*}

\section{Conclusions} \label{conclu}
In this paper, we presented the concept of activity disaggregation and discussed its potential advantages over traditional appliance load disaggregation. We developed a machine-learning-based framework to detect activities by leveraging different features, and validated the framework through a case study on house cooling/heating. We also discussed two approaches to activity modeling and showed numerical examples. 
For future work, we will extend the results for more activities and develop use cases to enable both utilities and consumers to benefit from activity detection and modeling techniques.

\bibliographystyle{IEEEtranTIE}
\bibliography{IEEEabrv.bib,ref.bib}

\end{document}